# Analysis and design of a SiGe-HBT based terahertz detector for imaging arrays applications

Hamed Ghodsi[1,*], Hassan Kaatuzian[1]

[1]*Photonics Research Laboratory, Electrical Engineering Department, Amirkabir University of Technology, Tehran 15914, Iran.*
*Corresponding author: hamed_88@aut.ac.ir*

**In this paper, we will introduce a new designed direct conversion terahertz detector and we will compare its characteristics with a prefabricated design in order to obtain better performance of our new proposed design. Differences between our new work and prefabricated device are in both physical structure by introducing an exponentially graded base SiGe-HBT instead of linearly graded ones, and configuration of components by means of merging two transistor of the initial common base structure into a single transistor with two base contacts, which equivalent to two transistors in common emitter configuration. New proposed detector, at first will be analyzed with a compact circuit model and then for more accurate analysis, two dimensional carrier transport analysis will be performed with numerical methods. Comparison of new and prefabricated detector will be done by simulation of both new and prefabricated devices with the same simulator. Because of only a slight change in transistor physical structure, if our simulation results and previous empirical data have a good matching for prefabricated device then with a high degree of precision we can claim our comparison is verified, but of course, for better confirmation of our ideas, fabrication and experimental measurements should be done in the next steps. Responsivity and minimum noise equivalent power of new detector are about 4.9A/W and 6.5pW/Hz1/2 respectively, while these characteristics for prefabricated detector, are about 1A/W and 50pW/Hz1/2 respectively. Also about 231μW/pixel decrement in power consumption for the same responsivity, and a same bandwidth have been achieved.**

## 1. Introduction

For many applications of terahertz waves, developing high precision, low noise, and low power compact detectors are seriously required. Some of these various applications are for example in medical imaging[1], spectroscopy[2] and sensing[9], where many biochemical molecules have strong spectral fingerprints at terahertz frequencies, and astronomical research[3], where a number of atomic and molecular emission lines in the THz range that are key diagnostic probes of the interstellar medium.

For this purpose, various methods and devices are used, where some of these methods can be found in [4, 5, 6, and 7]. In this paper silicon-germanium hetero junction, bipolar transistors in a direct conversion detector system are used. Direct conversion detector systems in comparison with traditional heterodyne detectors have less electrical components (more compact and appropriate for very large arrays) and thus better noise performance and lower power consumption while having good responsivity performance.[4, 7]

Si/SiGe HBTs displaced III–V HBTs in many applications mainly because of their supreme potential for large-scale integration and technological proximity to very mature Si processes.[8] In this work we are working on SiGe-HBTs (As shown in Figure.1) which are used in a direct conversion detector system. This direct conversion detector works based on nonlinearity of base-emitter junction diode, thus its operation is not limited by technology frequency limits ($f_T/f_{max}$) and can be used in frequencies over 650GHz in this work.

In our previous works, we started with analysis of optoelectronic mixers made from III-V HBT circuits in order to obtain higher frequency response [12], or higher conversion gain, using structural design alteration [11]. In addition, we focused on SiGe-HBT circuits [10, 13], to design Silicon compatible integrated circuits. In this study, in continuation of our recent works in SiGe-HBT, after explanations about detector design and implementation in section (2), in section(3), analysis of detector performance and theoretical viewpoints will be explained. Then in section(4), we'll introduce our ideas about the new detector design and in section(5), we are ready to compare our new proposed detector design (Refer to Figure.1 and Figure.7) with the initial detector implementation[7].There is also a conclusion section.

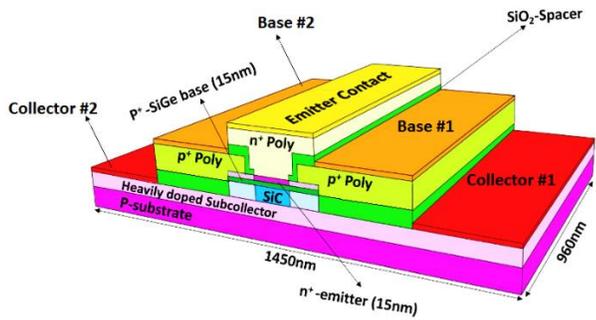

Fig. 1. 3D structure of transistor, used in this work. (Base #1 and Base #2 are not externally connected.)

## 2. Detector design

In direct conversion, detector systems the key element is a nonlinear device, which can generate second harmonic of, input terahertz signal in order to obtain square-law rectification. With a low-pass filter at the output, terahertz signal and its second and higher harmonics are filtered and as a result, output voltage and current will be proportional to terahertz input power. This nonlinear element is usually a diode, which also needs isolations between DC and RF signals. As a better implementation, two diodes in differential configuration are preferred because of creating an AC ground at the output that has an advantage for elimination of output RF isolation part.[7]

In a bipolar technology, a single diode implementation is done with a transistor by connecting its collector and base terminals. Nevertheless, there is a big disadvantage, because bipolar transistors are vertical devices and thus collector area is very large in comparison with the emitter. Therefore, performance of this detector will be effectively reduced by means of capacitive coupling to the substrate. This coupling effect wastes the terahertz input power and prevents it from being injected into the diode device.[7]

A solution to this problem is to use a differential common emitter configuration of two transistors as shown in Figure.2. In this configuration antenna is directly connected to the transistor base terminals and output current will be filtered for omitting terahertz frequencies and amplified with a precise low noise trans-impedance amplifier with nearly zero input impedance. In this work terahertz input signal is modulated with a 125 KHz signal in order to decrease low frequency noises and temperature dependent offset of bias circuit.

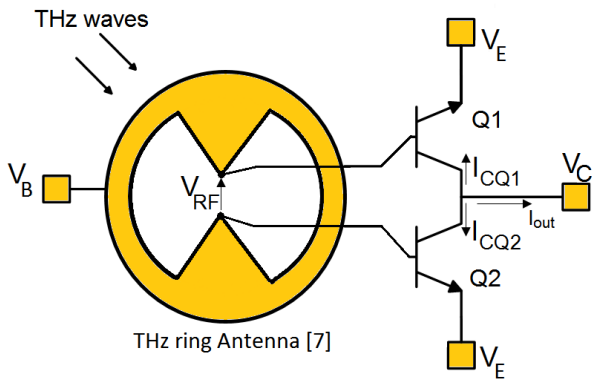

Fig. 2. Antenna coupled Common-emitter direct conversion detector of this work

A possible implementation of a 3×5 array of detectors, as shown in Figure.3 will be covered by a silicon lens which is diameter about 3 mm and thickness about 1.88 mm which is constructed to converge the input THz wave into the chip with a Gaussian profile. Further details about detector implementation, transistor fabrication and experimental setup for empirical studies, can be found in [7, 10, and 15].

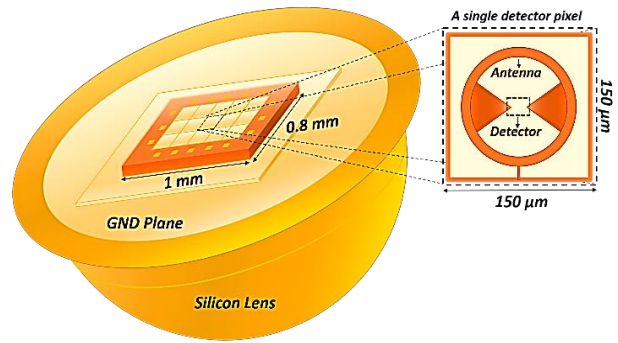

Fig. 3. Physical design of overall detector system with the silicon lens

## 3. Theoretical Analysis of detector performance

### A. Defining detector performance criteria:

Performance of terahertz detectors can be compared in several aspects but responsivity, noise equivalent power, power consumption and bandwidth are the most important characteristics. Optical responsivity is a measure of the electrical output strength per optical input power, which can be measured in current mode ($R_I$) or voltage mode ($R_V$). An accurate measurement of the optical responsivity requires an accurate measurement of the available optical input power ($P_{in}$) to the detector. A terahertz source with a measured total output power ($P_{TX}$) and a specified antenna gain ($G_{TX}$) is used to illuminate the detector placed at a distance ($r$). The available input power to the detector according to Friis transmission equation [14] is then given by: [7]

$$P_{in} = \frac{P_{TX} G_{TX}}{4\pi r^2} A_{eff} \qquad (1)$$

Where $A_{eff} = D_{RX}\lambda^2/(4\pi)$, where $\lambda$, is the wavelength and ($D_{RX}$) is receiver antenna diversity. Noise equivalent power (NEP) can be calculated from ratio of the spot noise at the output in a specific modulation frequency and output responsivity:

$$NEP = \frac{I_n}{R_I} \qquad (2)$$

Power consumption is also measured from product of collector bias voltage and its current and finally bandwidth in this work is defined to be the point of 3dB decrement in responsivity from responsivity in 650GHz.

### B. Analysis by compact circuit modeling of transistor

At first, we will try to have an overview to the most effective parameters of device physical structure on detector performance by means of a simplified compact circuit modeling (Refer to Figure.4). Of course, this model has lot of inaccuracies but this is not very important because for accurate analysis we will use two dimensional carrier transport equations.

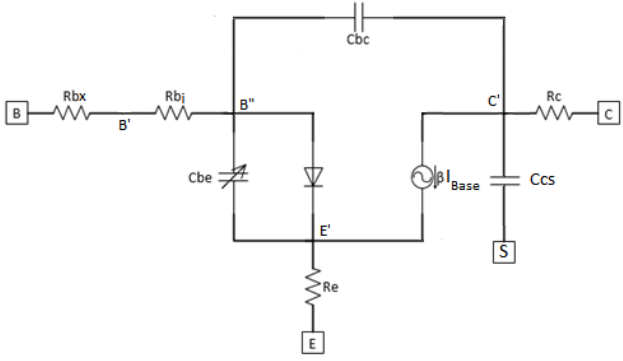

Fig. 4. Simplified compact transistor model for our initial analysis.

For simplicity, we divide our analysis into two parts. At first small signal analysis (similar to Hybrid-Pi model) will be performed to obtain ratio of voltage drop across base-emitter diode ($V_B''$-$V_E'$) to input terahertz voltage then by means of Tailor series expansion of diode exponential voltage-current expression we will calculate ratio of output current in modulation frequency to input terahertz power which means detector responsivity. From the circuit model, we can write (3) in small signal regime and for terahertz signal:

$$\begin{bmatrix} 1 & -\frac{j\omega r_\pi C_{be} R_B + R_B}{r_\pi + R_B + j\omega R_B r_\pi (C_{be}+C_{bc})} & \frac{j\omega r_\pi C_{bc} R_B}{r_\pi + R_B + j\omega R_B r_\pi (C_{be}+C_{bc})} \\ 1 & -\frac{r_\pi + (1+g_m r_\pi)R_E + j\omega C_{be} R_E r_\pi}{(1+g_m r_\pi)R_E + j\omega} & 0 \\ g_m R_C - j\omega C_{bc} R_C & -g_m R_C & j\omega C_{bc} R_C + j\omega C_{cs} + 1 \end{bmatrix}$$

$$\times \begin{bmatrix} V_B'' \\ V_C' \\ V_E' \end{bmatrix} = \begin{bmatrix} -\frac{\frac{v_{THz}}{2} r_\pi}{r_\pi + R_B + j\omega R_B r_\pi (C_{be}+C_{bc})} \\ 0 \\ 0 \end{bmatrix}$$

(3)

In order to have a better responsivity, ratio of voltage drop across base emitter diode to the input terahertz voltage should be larger and in ideal case should be one. From experimental measurements of transistor parasitic elements in zero bias condition [15] and because of several aggressive scaling steps, resistive elements and junction capacitances are lowered as much as possible. Also because of theoretical basis of this work and without any fabrication and experimental measurements, we prefer not to apply more scaling steps for further decrement of transistor parasitic elements. However, because base-emitter junction capacitance has much larger value than the other capacitances and it is in parallel with base-emitter diode and creates an undesired current path to ground for wasting large part of the input terahertz signal, it has an important effect on value of effective terahertz voltage drop on base-emitter diode. Base-emitter capacitance in a SiGe-HBT as shown in Figure.5 has several components:

$$C_{BE} = C_{BE_{OX}} + C_{BE_i}$$
$$C_{BE_i} = C_{BE_{junction}} + C_{BE_{diff}}$$
(4)

Where SiO$_2$ spacer capacitances are modeled in $C_{BEox}$ and junction capacitance and diffusion capacitance is also modeled in $C_{BEi}$. Spacer capacitances are related to geometry of spacers, which is dependent to fabrication process and because of complex geometry, derivation of a compact equation, is not straightforward. However, because of being much smaller than the other components, spacer capacitances can be neglected in our overall analysis. Junction and diffusion capacitances are the most important parts of overall base-emitter capacitance as mentioned in (5) and (6) respectively, but in many cases diffusion capacitance is dominant.[16]

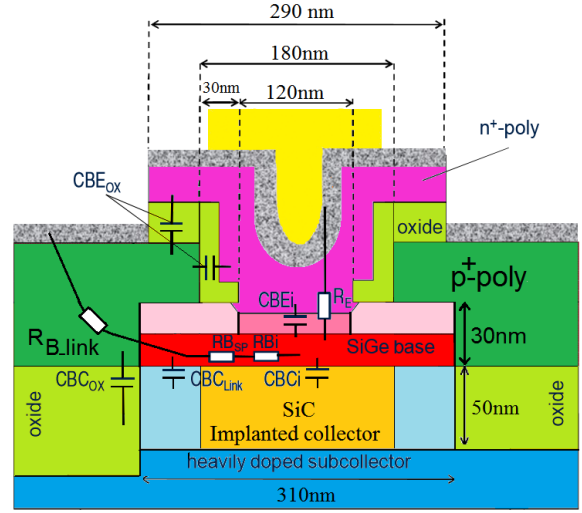

Fig. 5. Parasitic components in physical structure of SiGe-HBT transistor of this work. [15]

$$C_{BEj} = A_E \sqrt{\frac{q\varepsilon_s N_{AB} \cdot N_{DE}}{2(N_{AB}+N_{DE})(V_{bi(BE)} - V_a)}}$$ (5)

Where $N_{DE}$ and $N_{AB}$ are average doping levels of emitter and base respectively and $A_E$ is emitter effective area. In addition, $V_a$ is applied base-emitter voltage and $V_{bi(BE)}$ is junction built in voltage.

$$C_{BEdiff} = \frac{I_s e^{\frac{V_{BE}}{V_T}}}{V_T} \tau_B$$ (6)

Where "$I_s$" is reverse saturation current and "$\tau_B$" is base transit time and will be further explained later in this paper. As mentioned before decreasing capacitances with modification of device geometry and doping levels isn't our purpose, so we'll try to decrease base transit time which can effectively decreases diffusion capacitance also overall base-emitter capacitance. Also with smaller base transit time cut off frequency and bandwidth of detector can be effectively improved without increment in collector current and thus power consumption and shot noise levels.

Now with knowledge about effective value of terahertz voltage drop across base emitter diode by means of solving matrix equations of (3), we can focus on exponential characteristics of diode for calculating desired output current:

$$I_{diode} = \frac{I_s}{\beta} e^{\frac{qv_{BE}}{kT}} = \frac{I_s}{\beta} e^{\frac{qV_{BE}+v_{be}}{kT}} =$$

$$\frac{I_s}{\beta} e^{\frac{qV_{BE}}{kT}} e^{\frac{qv_{be}}{kT}} = \frac{I_{C(bias)}}{\beta} e^{\frac{q\zeta v_{THz} \sin(\omega_{mod} t)\sin(\omega_{THz} t)}{kT}}$$

$\rightarrow$

$$\rightarrow I_{diode} = \frac{I_{C(bias)}}{\beta} \left( 1 + \frac{q\zeta v_{THz} \sin(\omega_{mod}t)\sin(\omega_{THz}t)}{kT} + \frac{\left(\frac{q\zeta v_{THz} \sin(\omega_{mod}t)\sin(\omega_{THz}t)}{kT}\right)^2}{2} + O(x^3) \right) \quad (7)$$

Where $\zeta$ is included for effects of transistor parasitic effects on voltage drop across base emitter diode according to (3). Desired output current is caused by square component so we focus only on this component and thus we have:

$$\text{square component: } \beta \frac{I_{C(bias)}}{\beta} \frac{\left(\frac{q\zeta v_{THz} \sin(\omega_{mod}t)\sin(\omega_{THz}t)}{kT}\right)^2}{2}$$

$$= I_{C(bias)} \frac{\left(\frac{q\zeta v_{THz} \sin(\omega_{mod}t)}{kT}\right)^2}{4} (1 - \cos(2\omega_{THz}t)) \quad (8)$$

Amplitude of desired output current is: $I_{C(bias)} \left(\frac{q\zeta v_{THz}}{2kT}\right)^2$

As can be seen from (8) Responsivity is effectively dependent on $\zeta$, which has a complicated inverse dependency on bias current (capacitive coupling) due to (3) and thus an optimum bias point for maximum responsivity can be found.

### C. Carrier transport equations for more accurate analysis

For precise analysis of detector performance, we will try to solve set of nonlinear differential equations of carrier transport in a two-dimensional hetero structure system, using Newton numerical method. For this purpose, we should use current continuity equations for both holes and electrons, and Poisson equation like homogeneous case. In addition, changing dielectric constant should be taken into account and current density equations must be modified to take into account the non-uniform band structure. This procedure starts with current density expressions:[17]

$$\overrightarrow{J_n} = -\mu_n n \nabla \phi_n \quad , \quad \overrightarrow{J_p} = \mu_p p \nabla \phi_p \quad (9)$$

Where $\phi_n$ and $\phi_p$ are quasi-Fermi potentials:

$$\phi_n = \frac{1}{q} E_{FN} \quad , \quad \phi_p = \frac{1}{q} E_{FP} \quad (10)$$

Fermi energy levels are expressed in this form:

$$E_{FN} = E_C + kT \ln \frac{n}{N_C} - kT \ln \gamma_n n \quad \gamma_n = F_{1/2}^{-1}\left(\frac{n}{N_C}\right)$$

$$E_{FP} = E_V + kT \ln \frac{p}{N_V} - kT \ln \gamma_p p \quad \gamma_n = F_{1/2}^{-1}\left(\frac{p}{N_V}\right) \quad (11)$$

Where $\gamma_n$ and $\gamma_p$ are coefficients to involve Fermi-Dirac statistics. The conduction and valence band edge energies can be written as:

$$E_C = q(\psi_0 - \psi) - \chi$$
$$E_V = q(\psi_0 - \psi) - \chi - E_g \quad (12)$$

Where $\chi$ is position dependent electron affinity and $E_g$ is position dependent bandgap and $\psi_0$ is some reference potential, which can be selected in this form:

$$\psi_0 = \frac{\chi_r}{q} + \frac{kT}{q} \ln \frac{N_{C_r}}{n_{i_r}} = \frac{\chi_r + E_g}{q} - \frac{kT}{q} \ln \frac{N_{V_r}}{n_{i_r}} \quad (13)$$

Where "r" index indicates that parameters are taken from an arbitrary reference material. By combining equations, (9) to (13) current densities can be obtained in general case as:

$$\overrightarrow{J_n} = kT\mu_n \nabla n - q\mu_n n \nabla \left( \psi + \frac{kT}{q} \ln \gamma_n + \frac{\chi}{q} + \frac{kT}{q} \ln \frac{N_C}{n_{i_r}} \right)$$

$$\overrightarrow{J_p} = kT\mu_p \nabla p - q\mu_p p \nabla \left( \psi + \frac{kT}{q} \ln \gamma_p + \frac{\chi + E_g}{q} + \frac{kT}{q} \ln \frac{N_V}{n_{i_r}} \right) \quad (14)$$

The above equations cannot describe carrier transport behavior completely because at the interface of two different semiconductors mainly in abrupt hetero-junctions thermionic emission and field emission transport (tunneling) equations should be taken into account as follows:

$$\overrightarrow{J_n} = qv_n(1+\delta)\left(n^+ - n^- \exp\left(\frac{-\Delta E_C}{kT}\right)\right)$$

$$\overrightarrow{J_p} = -qv_p(1+\delta)\left(p^+ - p^- \exp\left(\frac{-\Delta E_V}{kT}\right)\right) \quad (15)$$

Where $J_n$ and $J_p$ are electron and hole current densities from the "-"region to "+". $v_n$ and $v_p$ are electron and hole thermal velocities. $\Delta E_C$ and $\Delta E_V$ are conduction band and valence band energy changes going from the "-"region to "+" respectively. $\delta$ parameter represents the contribution of thermionic field emission (tunneling) which is defined in (16).

$$\delta = \frac{1}{kT} \int_{E_{min}}^{E_C^+} A \times \exp\left(\frac{E_C^+ - E_X}{kT}\right) dE_X$$

$$A = \exp\left(\frac{-4\pi}{h} \int_0^{X_E} \left[2m_n^*(E_C - E_X)\right]^{\frac{1}{2}} dx\right) \quad (16)$$

Where $E_x$ is energy component in x direction and $X_E$ is corresponds to $E_x$ and $E_{min}$=max {$E_C^{0-}$, $E_C^{\infty}$}.

### D. New proposed ideas

As mentioned before for improvement of detector performance we should decrease capacitive parasitic effects in base-emitter junction of the transistor. We will do this by means of deceasing base transit time and thus diffusion capacitance. To do so, we will introduce a new exponentially graded alloyed base. For explanation of our idea, we start with a one-dimensional simplified equation: [18]

$$I_C = -qA_E D_n \left[ \frac{dn(x)}{dx} - \frac{q\varepsilon_B}{kT} n(x) \right] \quad (17)$$

Where $I_C$ is collector current and $n(x)$ is minority electron concentration in the base region, $A_E$ is emitter area and $\mu_n$ and $D_n$ are minority electron mobility and diffusion constant in the base region respectively and $\varepsilon_B$ is internal field. We can derive minority electron concentration in the base region from (17) as:

$$n(x) = \frac{I_C}{qA_E D_n} \frac{kT}{q\varepsilon_B} \left\{ 1 - \exp\left[ -\frac{q\varepsilon_B}{kT}(X_B - x) \right] \right\} \quad (18)$$

Where $\varepsilon_B$ is the base internal field due to electron affinity gradual changes in the base region. Although in reality, internal field isn't constant but in case of relatively large internal fields and with constant current in the base region, from (18) it can be seen at the emitter side of base, gradient of minority electrons are relatively small and main part of current is due to drift component and internal field and at the collector side because of nearly zero electron concentration diffusion current is dominant. Thus with constant current level assumption and large internal field we can assume a constant internal field equal to slope of bandgap changes at the emitter side in the whole base region and after some mathematical simplifications base transit time can be written as:

$$\tau_B = \frac{Q_B}{I_C} = \frac{A_E \int_0^{X_B} n(x)dx}{I_C} = \frac{X_B^2}{2D_n} f(\kappa) \quad (19)$$

$$f(\kappa) = \frac{2}{\kappa}\left(1 - \frac{1}{\kappa} + \frac{1}{\kappa}e^{-\kappa}\right)$$

Where $X_B$ is base thickness, $D_n$ is minority carrier diffusion constant and $\kappa$ is calculated from:

$$\kappa = \frac{q\varepsilon_B}{kT} X_B = \frac{X_B}{kT} \frac{dE_{g(base)}}{dx}\bigg|_{x=0} \quad (20)$$

Bandgap of SiGe alloys is related to Ge percentage by this equation: [19]

$$E_g = 1.155 - 0.874x + 0.376x^2 \quad (21)$$

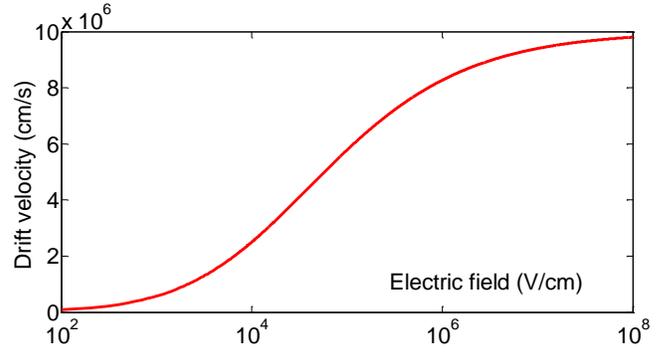

Fig. 6. Drift velocity Vs Electric field in SiGe base for $N_A=10^{19} cm^{-3}$. Extracted from [17] and [20]

As mentioned in [15] initial transistor with a 15nm base has a linearly graded base from 15% at the emitter side to 28% at the collector side thus maximum internal field intensity is about $6\times10^4$ V/cm. With respect to Figure.6 internal field can be further increased before velocity saturation occurs, but for decreasing base transit time with increasing internal field, critical SiGe base thickness should also be considered. As mentioned in [21] for a 15nm base, maximum Ge percentage is about 30% and thus increasing internal field with increasing Ge percentage is not possible in our case. But in order to have a large internal field, with the same maximum Germanium percentage we can use an exponential grading profile as mentioned in (22) and shown in Figure.7.

$$Ge\ percentage = 28.67 - 13.67 e^{\frac{-x\ (nm)}{5nm}} \quad (22)$$

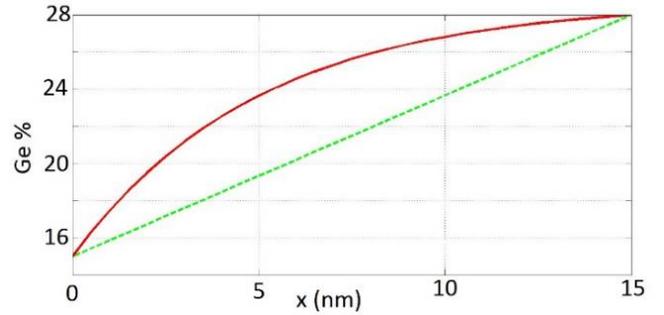

Fig. 7. New proposed Ge exponential grading Vs initial linear grading profile

Using this exponential Germanium profile and (17) to (22), base internal field will be about $1.97\times10^5$ V/cm, which is 3.3 times larger than the initial value and thus base transit time will be deceased about 60%. However, is it possible to fabricate this transistor? Of course, for exact answer to this question, this transistor should be fabricated and experimental measurements should be done, but we can predict possibility of this idea. So far, for SiGe base deposition, several methods like LPCVD [22] and, UHV/CVD [23] have been developed. For deposition of the proposed exponentially graded layer, the most important concern is about possibility of forming a grading profile 3.15 times sharper than the initial transistor of [15] at the emitter side of base region. Moreover, if this could be possible, in the rest of base region with smaller grading slopes; formation of base profile is also possible. With respect to experimentally reported transistor of [22] with a nearly abrupt Ge profile in the LPCVD grown base region, it seems that formation of our proposed grading profile is also possible. Improvements of this idea on transistor characteristics are shown in Table I. initial and new transistor is compared.

**Table I. Comparison between new and initial transistor**

| Initial transistor (simulation) | | | New transistor (simulation) | | |
|---|---|---|---|---|---|
| β | $f_T$ | $f_{max}$ | β | $f_T$ | $f_{max}$ |
| 698 | 265GHz | 425GHz | 975 | 370GHz | 705GHz |

Because of several scaling steps and very small parasitic resistances about few ohms [15] thermal noise of these resistances are very small and the dominant noise is shot noise. So average noise current in collector which is used for calculation of noise equivalent power (NEP) of detector is about $2qI_C$ for each transistor and $4qI_C$ for two transistors in common emitter configuration. To have a lower NEP and lower power consumption we should decrease bias current but without any reduction in responsivity of the detector and this cannot be possible. To overcome this problem we propose to use two base contacts for our transistor and using these base contacts as base terminals of initial common emitter design. Note that these two contacts cannot be externally been contacted because they are not for one transistor and in the other word there are two transistors but merged into one device. This configuration gives us several advantages over the initial design including less parasitic elements and thus higher responsivity and bandwidth, lower noise levels and also lower power consumption because of omitting one transistor. In Figure.8, our final design is shown.

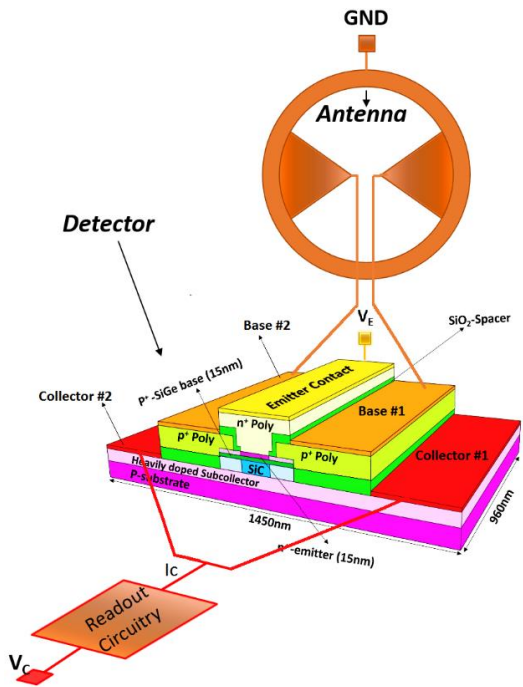

Fig. 8. New Proposed detector configuration

## 4. Results and discussion

Comparison of new and prefabricated detector is done by simulating both of them with circuit modeling and carrier transport approaches. Although result of both methods have a good matching with each other but to be more accurate only results of carrier transport analysis is shown here. Because of only a slight change in transistor physical structure, if our simulation results and previous empirical data have a good matching for prefabricated device then with a high degree of precision we can claim our comparison is correct, but of course, for better confirmation of our ideas, fabrication and experimental measurements should be done in the next steps. As an example in Figure.9, there is a good matching between experiments and our simulations for prefabricated device.[10]

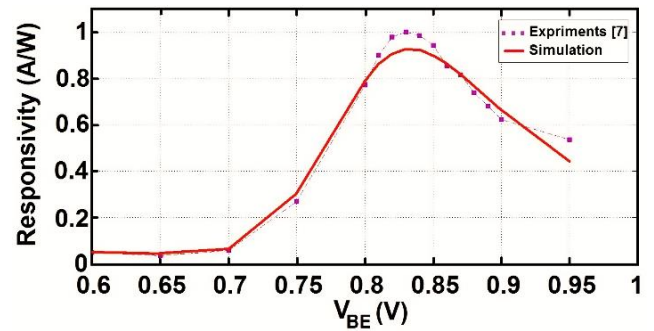

Fig. 9. Simulation Vs experimental data about Responsivity of prefabricated detector for optimum $V_{CE}$

Measurements, which we used for comparison between new proposed detector and prefabricated detector, as mentioned before, are responsivity, noise equivalent power, and bandwidth and power consumption. In Figure.10 to Figure.12 and finally in TABLE II performance characteristics of new designed device and prefabricated detector device can be compared.

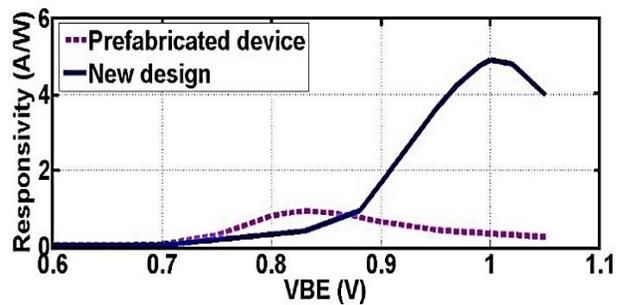

(a)

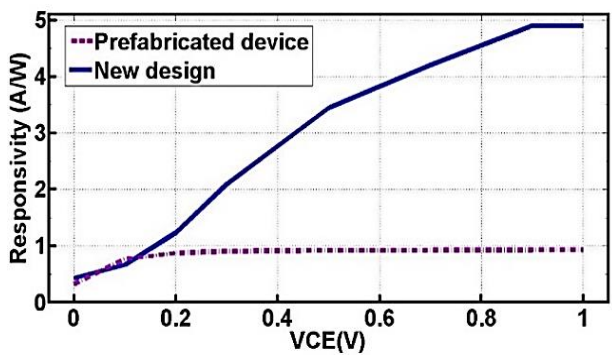

(b)

Fig. 10. (a) Responsivity of New designed Vs prefabricated detector data for optimum $V_{CE}$. (b) Responsivity of New designed Vs prefabricated detector data for optimum $V_{BE}$

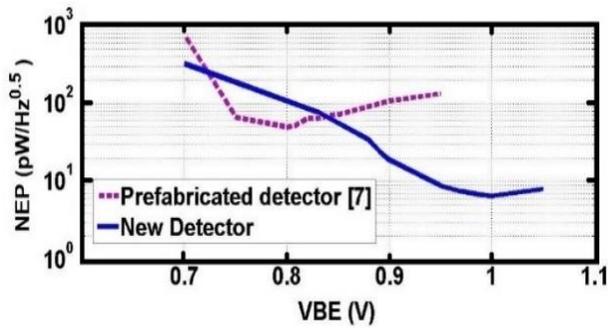

(a)

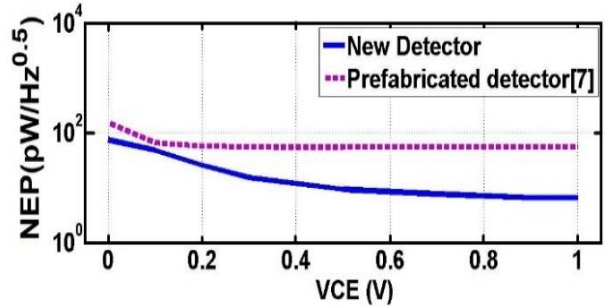

(b)

Fig. 11. (a) Noise equivalent power of New designed Vs prefabricated detector data for optimum $V_{CE}$. (b) Noise equivalent power of New designed Vs prefabricated detector data for optimum $V_{BE}$

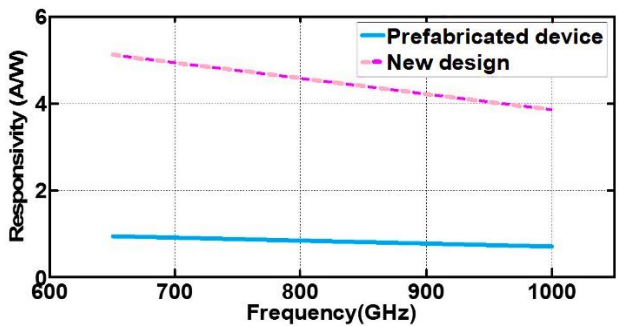

Fig. 12. Frequency response of New designed Vs prefabricated detector data for optimum bias point

**Table II. Comparison between characteristics of new designed and prefabricated detectors**

|   | Detector of [7] (Experiments) | Detector of [7] (simulation) | New Detector (simulation) |
|---|---|---|---|
| Responsivity | 1 A/W | 0.927 A/W | 4.9 A/W |
| Noise equivalent power | 50 pW.Hz$^{-1/2}$ | 50 pW.Hz$^{-1/2}$ | 6.5 pW.Hz$^{-1/2}$ |
| 3dB Bandwidth | (0.65-0.85) THz | (0.65-1.37) THz | (0.65-1.37) THz |
| Power Consumption | 331 µW/pixel $R_I$=1A/W | 331 µW/pixel $R_I$=0.93A/W | 100 µW/pixel $R_I$=1.12A/W |

## 5. Conclusion

In this paper, we have designed a new direct conversion detector for terahertz frequencies by mean of analyzing performance of a prefabricated detector system and applying effective modifications in both physical structure of transistors and configuration of components. Using exponentially graded base transistors and a single transistor with two base contacts which works like two transistors in differential common emitter configuration, several major improvements in detector performance characteristics were achieved including Responsivity improvement about **490%** with the same bandwidth and minimum noise equivalent power (NEP) at detector output is decreased about **87%** and finally power consumption per pixel for the same responsivity is decreased about **70%**. However, fabrication of transistor will be more complex.

Next steps of this research can be implementation of the new designed device in order to confirm theoretical results of this paper by an empirical approach and use of this detector in larger arrays in one of the mentioned applications like a closed box terahertz scanning, similar to [7]. Also about performance optimization, working on Si lens, antenna and further transistor improvement by means of more scaling steps and higher internal field can be suggested.